\documentclass[aps,twocolumn,showpacs,byrevtex,prl,reprint,nofootinbib]{revtex4-1}
\usepackage{xspace}
\usepackage{graphicx}
\usepackage{dcolumn}
\usepackage{bm}
\usepackage{rotating}
\usepackage{color}
\usepackage{verbatim} 
\usepackage{multirow}
\usepackage{epstopdf}

\newcommand{\un}[1]{\ensuremath{\,\mathrm{#1}}}

\def \mevcc {\ensuremath{\un{MeV}/c^2}\xspace}

\def \gevcc {\ensuremath{\un{GeV}/c^2}\xspace}

\def \ks {K_{S}}
\def \ee {e^+e^-}
\def \psip {\psi(3686)}
\def \jp {J/\psi}
\def \chict {\chi_{c2}}

\def \kk {K^+K^-}

\begin{document}
\title{\boldmath Observation of the helicity-selection-rule suppressed decay of the $\chi_{c2}$ charmonium state }

\author{
  \begin{small}
    \begin{center}
      M.~Ablikim$^{1}$, M.~N.~Achasov$^{9,e}$, S. ~Ahmed$^{14}$,
      X.~C.~Ai$^{1}$, O.~Albayrak$^{5}$, M.~Albrecht$^{4}$,
      D.~J.~Ambrose$^{44}$, A.~Amoroso$^{49A,49C}$, F.~F.~An$^{1}$,
      Q.~An$^{46,a}$
% ...
      \\
      \vspace{0.2cm}
      (BESIII Collaboration)\\
      \vspace{0.2cm} {\it
        $^{1}$ Institute of High Energy Physics, Beijing 100049, People's Republic of China\\
% ...
$^{4}$ Bochum Ruhr-University, D-44780 Bochum, Germany\\
        $^{5}$ Carnegie Mellon University, Pittsburgh, Pennsylvania 15213, USA\\
% ...
$^{9}$ G.I. Budker Institute of Nuclear Physics SB RAS (BINP), Novosibirsk 630090, Russia\\
% ...
$^{14}$ Helmholtz Institute Mainz, Johann-Joachim-Becher-Weg 45, D-55099 Mainz, Germany\\
% ...
$^{44}$ University of Rochester, Rochester, New York 14627, USA\\
% ...
$^{46}$ University of Science and Technology of China, Hefei 230026, People's Republic of China\\
% ...
$^{49}$ (A)University of Turin, I-10125, Turin, Italy; (B)University of Eastern Piedmont, I-15121, Alessandria, Italy; (C)INFN, I-10125, Turin, Italy\\
% ...
$^{a}$ Also at State Key Laboratory of Particle Detection and Electronics, Beijing 100049, Hefei 230026, People's Republic of China\\
% ...
$^{e}$ Also at the Novosibirsk State University, Novosibirsk, 630090, Russia\\
% ...
      }
    \end{center}
    \vspace{0.4cm}
  \end{small}
}

\affiliation{}


\begin{abstract}
 The decays of $\chi_{c2}\to K^{+}K^{-}\pi^{0}$, $\ks K^{\pm}\pi^{\mp}$ and $\pi^{+}\pi^{-}\pi^{0}$ are studied with the $\psip$ data samples collected with the Beijing Spectrometer (BESIII). For the first time, the branching fractions of $\chi_{c2}\to K^{\ast}\overline{K}$, $\chi_{c2}\to a_{2}^{\pm}(1320)\pi^{\mp}/a_{2}^{0}(1320)\pi^{0}$ and $\chi_{c2}\to \rho(770)^{\pm}\pi^{\mp}$ are measured. Here $K^{\ast}\overline{K}$  denotes both $K^{\ast\pm}K^{\mp}$ and its isospin-conjugated process $K^{\ast 0}\overline{K}{}^{0}+\rm {c.c.}$, and $K^{\ast}$ denotes the resonances $K^{\ast}(892)$, $K^{\ast}_{2}(1430)$ and $K^{\ast}_{3}(1780)$. The observations indicate a strong violation of the helicity selection rule in $\chi_{c2}$ decays into vector and pseudoscalar meson pairs. The measured branching fractions of $\chi_{c2}\to K^{\ast}(892)\overline{K}$ are more than 10 times larger than the upper limit of $\chi_{c2}\to \rho(770)^{\pm}\pi^{\mp}$, which is so far the first direct observation of a significant $U$-spin symmetry breaking effect in charmonium decays.
\end{abstract}

\maketitle
The helicity selection rule (HSR)~\cite{Brodsky:1981kj,Chernyak:1981zz,Chernyak:1983ej} is one of the most important consequences of perturbative quantum chromodynamics  (pQCD) at leading twist accuracy. In the charmonium energy region, although there are observations that pQCD plays a dominant role, there are also many hints that non-perturbative mechanisms can become important~\cite{Chernyak:1983ej,nonpqcd1,nonpqcd2,nonpqcd3}. Exclusive decays of the $P$-wave charmonium state $\chi_{c2}\to VP$, where $V$ and $P$ denote light vector and pseudoscalar mesons, respectively, are ideal for testing the HSR and pinning down the mechanisms that may violate the leading pQCD approximation.

Another reason the decays of $\chi_{c2}\to VP$ are of great interest is that this process is ideal for probing the long-range interactions arising from intermediate $D$-meson loop transitions. As pointed out in Ref.~\cite{zhaoq}, if the intermediate $D$-meson loops provide the non-perturbative mechanism to violate the HSR, this can be identified by the measurements of $\chi_{c2}\to K^{\ast}(892)\overline{K}$  and $\chi_{c2}\to \rho(770)^{\pm}\pi^{\mp}$.

In this Letter, we present a partial wave analysis (PWA) of the process $\chi_{c2}\to K\overline{K}\pi$ (denotes $K^{+}K^{-}\pi^{0}$ and $\ks K^{\pm}\pi^{\mp}$) and a measurement of $\chi_{c2}\to \pi^{+}\pi^{-}\pi^{0}$. We have two $\psip$ samples of $(106.8\pm 0.8)\times 10^{6}$ (160~pb$^{-1}$)~\cite{wang} and $(341.1\pm 2.1)\times 10^{6}$ (510~pb$^{-1}$)~\cite{wang1} events collected in 2009 and 2012 by BESIII~\cite{bes3}, respectively. Only the 2009 data sample is used in the analysis of $\chi_{c2}\to K\overline{K}\pi$, and the full data sample is used in  $\chi_{c2}\to\pi^{+}\pi^{-}\pi^{0}$  since it has a smaller branching fraction. An independent sample of about 44~pb$^{-1}$ taken at $\sqrt{s}=3.65$~GeV is utilized to investigate the potential background from the continuum process.  A sample of Monte Carlo (MC) simulated events of generic $\psip$ decays (inclusive MC sample) is used to study backgrounds. The optimization of the event selection and the estimation of physics backgrounds are performed with Monte Carlo simulations of $\psip$ inclusive/exclusive decays.

The $\chi_{c2}$ candidates, produced in $\psip$ radiative decays, are reconstructed from the final states $K^{+}K^{-}\pi^{0}$, $\ks K^{\pm}\pi^{\mp}$, and $\pi^{+}\pi^{-}\pi^{0}$. Each charged track is required to have a polar angle $\theta$ in the main drift chamber (MDC) that satisfies $|\cos\theta|<0.93$, and have the point of closest approach to the $\ee$ interaction point within 10 cm in the beam direction ($|V_{z}|$), and 1 cm in the plane perpendicular to the beam direction ($V_{r}$). The energy loss $dE/dx$ in the MDC and the information from the time-of-flight (TOF) system are combined to form particle identification (PID) confidence levels (C.L.) for the $\pi,~K$, and $p$ hypotheses, and each track is assigned with the hypothesis corresponding to the highest C.L.
The $\ks$ candidates are reconstructed from two oppositely charged tracks with loose vertex requirements ($|V_{z}|<30$~cm and $V_{r}<10$~cm) and without PID (assumed to be pions). Then the candidate with invariant mass closest to the $\ks$ nominal mass and the decay length provided by a secondary vertex fit algorithm greater than 0.25~cm, is selected for further study in the decay $\psip\to\gamma \ks K^{\pm}\pi^{\mp}$. The candidate events are required to have two charged tracks with zero net charge, where the tracks from the $\ks$ candidate are not taken into account.  Two pions and one kaon are required for the decays $\psip\to\gamma \pi^{+}\pi^{-}\pi^{0}$ and $\psip\to \gamma\ks K^{\pm}\pi^{\mp}$, respectively, and no PID requirement is applied  for the decay $\psip\to\gamma K^{+}K^{-}\pi^{0}$.
The photon candidates are required to have energy larger than 25 (50)~MeV in the Electromagnetic Calorimeter (EMC) barrel (end cap) region $|\cos\theta|<0.8$ ($0.86<|\cos\theta|<0.93$),
and have an angle relative to the nearest charged tracks larger than $10^\circ$. To suppress electronic noise and energy deposits unrelated to the event, the EMC cluster time must be within 700~ns  from the event start time. At least three and one photons are required for the decay $\psip\to\gamma \pi^{+}\pi^{-}\pi^{0}$/$K^{+}K^{-}\pi^{0}$ and $\psip\to\gamma \ks K^{\pm}\pi^{\mp}$, respectively.

 A fit with four kinematic constraints (4C) enforcing four-momentum conservation between the initial $\psip$ and the final state is performed for each process. If there are more photons than required in one event, all possible combinations of photons are considered  and only the one with the least $\chi_{4C}^{2}$ of the kinematic fit is retained for further analysis. The $\chi_{4C}^{2}$ is required to be less than 80 and 60 for the decay $\psip\to\gamma K^{+}K^{-}\pi^{0}$ and $\psip\to\gamma \ks K^{\pm} \pi^{\mp}$, respectively. The $\pi^{0}$ candidate is reconstructed from the two selected photons whose invariant mass is closest to the $\pi^{0}$ nominal mass, and satisfies $|M_{\gamma\gamma}-M_{\pi^{0}}|<10~\mevcc$. For the decay mode $\psip\to\gamma \pi^{+}\pi^{-}\pi^{0}$, a 5C kinematic fit is performed with an additional $\pi^{0}$ mass constraint, and $\chi^{2}_{5C}<60$ is required. To remove the backgrounds $\psip\to \pi^{0}\pi^{0}\jp$ ($\jp\to l^+l^-$, $l=e,\mu$), the invariant mass of $K^{+}K^{-}$/$\pi^{+}\pi^{-}$  is required to be less than 3.0~\gevcc for the decay $\psip\to \gamma K^{+}K^{-}\pi^{0}/\gamma \pi^{+}\pi^{-}\pi^{0}$. In the decay mode $\psip\to\gamma \pi^{+}\pi^{-}\pi^{0}$, the $\pi^{0}$ recoil mass is required to be less than 3.0~\gevcc to suppress the background $\psip\to\pi^{0}\jp$, and $M_{\gamma\pi^{0}}\not\in (0.7,0.85)~\gevcc$ is required to veto the background  $\psip\to \omega \pi^{+}\pi^{-}$ ($\omega\to\gamma \pi^{0}$).

\begin{figure}[htbp]
\includegraphics[width=8.2cm,height=7.0cm]{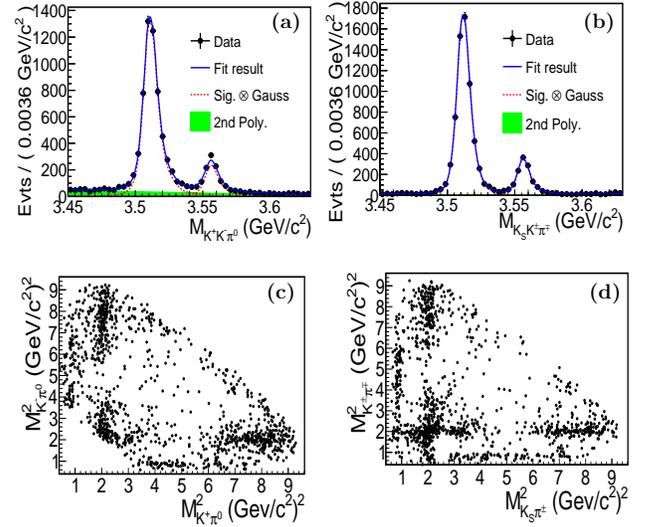}
\caption{(color online) Invariant mass distribution of (a) $K^{+}K^{-}\pi^{0}$ and (b) $\ks K^{\pm} \pi^{\mp}$ for the decay $\psip\to\gamma K\overline{K}\pi$, and the  corresponding Dalitz distributions (c) and (d) for the candidates within $|M_{KK\pi}-M_{\chi_{c2}}|<15$~\mevcc. The dots with error bars are for data, the blue solid curves are the overall fit results, the red dotted curves are the signals, and the green shaded areas are  the background. }
\label{gkkpi:2d}
\end{figure}

The $K\overline{K}\pi$ invariant mass for the decay $\psip\to\gamma K^{+}K^{-}\pi^{0}$ and $\psip\to\gamma\ks K^{\pm}\pi^{\mp}$ are shown in Fig.~\ref{gkkpi:2d}(a) and (b), respectively. The $\chi_{c1,2}$ signals appear prominently with a small background.  From the analysis of the $\psip$ inclusive MC sample and the continuum data at $\sqrt{s}=3.65$~GeV, the main backgrounds are from the decays $\psip\to \pi^0\pi^0 J/\psi$ ($J/\psi\to\pi^{+}\pi^{-}\pi^{0}/\mu\mu$), and $\psip\to K_{1}(1270)^{\pm}K^{\mp}$ ($K_{1}(1270)^{\pm}\to K^{\pm}\pi^{0}\pi^{0}/\rho(770)^{\pm} \ks$).
All of these backgrounds show a smooth distribution, and do not produce a peak around the $\chi_{cJ}$ mass region. Unbinned maximum likelihood fits are performed to the selected candidates, where the $\chi_{c1,2}$ signals are described with the MC simulated shapes convoluted with a Gaussian function accounting for the resolution difference between data and MC simulation, and the backgrounds are described with a $2^{nd}$ order  polynomial function. There are 1215 and 1176 candidate events  for $\chi_{c2}\to K^{+}K^{-}\pi^{0}$ and $\chi_{c2}\to \ks K^{\pm}\pi^{\mp}$ within the $\chi_{c2}$ signal region  $|M_{K\overline{K}\pi}-M_{\chi_{c2}}|\le 15~\mevcc$.
Non-$\pi^{0}$ ($K^{+}K^{-}\pi^{0}$ mode only) and non-$\chi_{c2}$ backgrounds are estimated with the events in the sideband regions, which are also used in the PWA as described in the following. The numbers of background events are estimated to be 240 and 80 for $\chi_{c2}\to K^{+}K^{-}\pi^{0}$ and $\chi_{c2}\to\ks K^{\pm} \pi^{\mp}$, respectively.

In the PWA, the process $\chi_{c2}\to K\overline{K}\pi$ is assumed to proceed via the quasi two-body decays, {\it i.e.} $\chi_{c2}\to a_2\pi$ and $K^*\overline{K}$ followed by $a_2\to K\overline{K} $ and $K^*\to K\pi$.  The amplitudes of the two-body decays are constructed with the helicity-covariant method~\cite{chung}. For a particle decaying into two-body final states, {\it i.e.} $A(J,m)\to B(s,\lambda) C(\sigma,\nu)$, where spin and helicity are indicated in the parentheses, its helicity covariant amplitude $F_{\lambda,\nu}$~\cite{chung} is:
\begin{equation}
\label{pwa:f}
 F_{\lambda,\nu}=\sum\limits_{LS}A g_{LS}\langle L0S\delta|J\delta\rangle\langle s\lambda\sigma-\nu|S\delta\rangle r^{L}B_{L}(r),
\end{equation}
where $A\equiv \sqrt{\frac{2L+1}{2J+1}}$, $g_{LS}$ is the coupling constant for the partial wave with orbital angular momentum $L$ and spin $S$ (with $z$-projection $\delta$), $r$ is the relative momentum between the two daughter particles in the initial particle rest frame, and $B_L$ is the barrier factor~\cite{zoubs:2003pj}. The conservation of parity is applied in the equation. Recent measurements show that the contributions of higher order magnetic and electric multipoles in the $\psip$ radiative transition to $\chi_{c2}$ are negligible, and the $E1$ transition is the dominant process~\cite{e1transition}. Hence, the helicity amplitudes are constructed to satisfy the $E1$ transition relation~\cite{E1relation}  and parity conservation, namely, $F_{1,2}=\sqrt{2}F_{1,1}=\sqrt{6}F_{1,0}$ and $F_{0,0}=0$. The corresponding $g_{LS}$ are taken as complex values. The relative magnitudes and phases are determined by an un-binned maximum likelihood fit to data with the package MINUIT~\cite{minuit}. The background contribution to the likelihood value is estimated with the events in the sideband regions and is subtracted~\cite{zhuc}. For the PWA method check, a input data is generated with inclusion of all states in the baseline solution, and coupling constants are fixed to the PWA solution. After the detector simulation and selection criteria, the same PWA fit procedure is performed, the fit results are consistent with that of the input data within the statistical errors.

As shown in the Dalitz plots of Fig.~\ref{gkkpi:2d}(c) and (d), clear signals for $K^{\ast}(892)$ and $K^{\ast}_{2}(1430)$  are observed in the $K\pi$ system. The resonances $K^{\ast}(892)$ and $K^{\ast}_{2}(1430)$ in the $K\pi$ system as well as the $a_{2}(1320)$  in the $K\overline{K}$ system, which has a significance larger than $8\sigma$ in both decay modes, are included in the baseline solution. For consistency, the $K_{3}^{\ast}(1780)$  signal, which is of a significance larger than $5\sigma$ in the decay $\chi_{c2}\to \ks K^{\pm}\pi^{\mp}$, but only $2\sigma$ in $\chi_{c2} \to K^{+}K^{-}\pi^{0}$, is also included. A contribution from the direct $\chi_{c2}\to K\overline{K}\pi$ three-body decay, which is parameterized with a non-resonant component with spin-parity $J^{p}=2^{+}$ in the $K\overline{K}$ system, is also considered. Other possible excited $K^{\ast}$ states in $K\pi$ system and the states in the $K\overline{K}$ system listed in PDG~\cite{PDG}, which have a significance less than $5\sigma$, are not included, but they are considered as a source of systematic uncertainty. The coupling constants for the charge-conjugate modes are treated to be the same.

\begin{figure}[hptb]
\includegraphics[width=8.2cm,height=3.5cm]{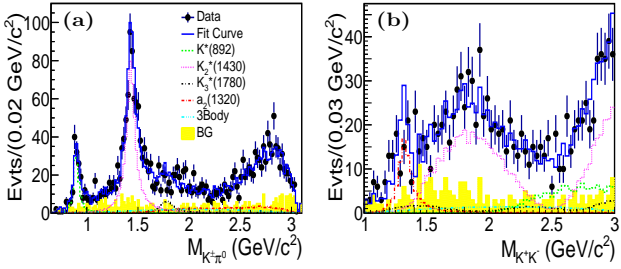}
\caption{(color online) Projections of the fit results onto the invariant mass of (a) $K^{\pm}\pi^{0}$ and (b) $K^{+}K^{-}$ in the decay $\chict\to K^{+}K^{-}\pi^{0}$,
where dots with error bars are for data, the blue solid histograms are the overall fit results, the yellow shaded histograms are for the background  estimated using $\chi_{c2}$ sideband events, and the contributions from different components are indicated in the inset.
\label{pwa:gkkpi0}}
\end{figure}

\begin{figure}[hptb]
\begin{center}
\includegraphics[width=8.2cm,height=7.0cm]{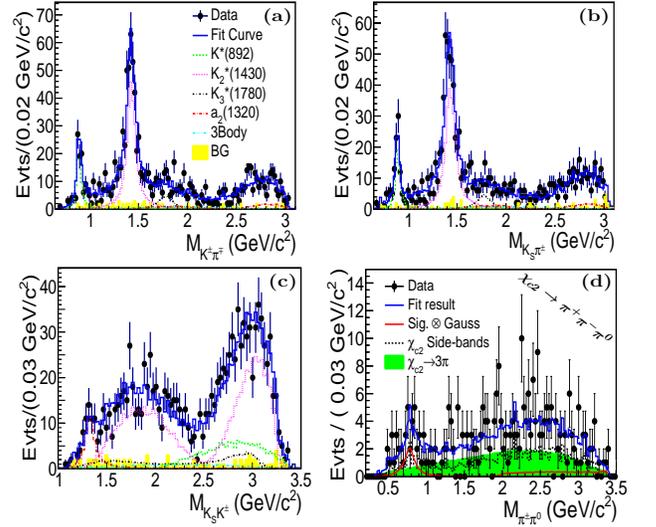}
\caption{(color online) Projections of the fit results onto the invariant mass of (a) $K^{\pm}\pi^{\mp}$, (b) $\ks\pi^{\pm}$ and (c) $\ks K^{\pm}$ in the decay $\chi_{c2}\to \ks K^{\pm}\pi^{\mp}$  as well as (d) $\pi^{\pm}\pi^{0}$ in the decay $\chi_{c2}\to \pi^{+}\pi^{-}\pi^{0}$,
where dots with error bars are for data, the blue solid histograms are the overall fit results, and the contributions from other components are indicated in the inset.}
\label{pwa:gkskpi}
\end{center}
\end{figure}

Figures~\ref{pwa:gkkpi0} and \ref{pwa:gkskpi}(a)-(c) show the invariant mass distribution and the projection of the PWA for the decays $\chi_{c2}\to K^{+}K^{-}\pi^{0}$ and $\chi_{c2}\to\ks K^{\pm}\pi^{\mp}$, respectively. The signal yields for the individual processes with a given intermediate state and the corresponding statistical uncertainties are calculated according to the fit results. The resultant branching fractions for the decays $\chi_{c2}\to K^{\ast}\overline{K}$ and $\chi_{c2}\to a_{2}(1320)\pi$ are summarized in Table~\ref{sumtab1}. The branching fractions for the processes including charged $K^{\ast}$ intermediate states are consistent between the two decay modes, and are combined by considering the correlation of uncertainties between the two modes~\cite{sunzt}. The $K^{\ast}$ isospin-conjugate modes are consistent with each other within $2\sigma$, as expected by isospin symmetry.

\begin{table}[hpbt]
\caption{The measured branching fractions $(\times 10^{-4})$ for the decays $\chi_{c2}\to (K\pi)\overline{K}$ and $\chi_{c2}\to (K\overline{K})\pi$. The first uncertainties are statistical, and the second are systematic (Here, $K^{\ast}$, $K^{\ast}_{2}$, $K^{\ast}_{3}$, and $a_{2}$ refer to $K^{\ast}(892)$, $K^{\ast}_{2}(1430)$, $K^{\ast}_{3}(1780)$, and $a_{2}(1320)$, respectively).
}
\label{sumtab1}
\begin{tabular}{lccc}
\hline\hline
 Mode& $K^{+} K^{-} \pi^0$ & $\ks K^\pm \pi^\mp$ & Combined \\ \hline
 $K^{\ast\pm}K^{\mp}$       &$1.8\pm 0.2\pm 0.2$ & $1.4\pm 0.2\pm0.2$ &$1.5\pm 0.1 \pm 0.2$ \\
 $K^{\ast 0}\overline{K}{}^0$ &$-$ & $1.3\pm 0.2\pm 0.2$ & $-$ \\
 $K_{2}^{\ast\pm}K^{\mp}$  &$18.2\pm 0.8\pm 1.6$ & $13.6\pm 0.8\pm 1.4$ & $15.5\pm 0.6\pm 1.2$ \\
 $K_{2}^{\ast 0}\overline{K}{}^0$&$-$ & $13.0\pm 1.0\pm 1.5$&$-$                         \\
 $K_{3}^{\ast\pm}K^{\mp}$  &$5.3\pm 0.5\pm 0.9$ & $5.9\pm 1.1\pm 1.5$ &$5.4\pm0.5\pm 0.7$ \\
 $K_{3}^{\ast 0}\overline{K}{}^0$ &$-$ & $5.9\pm 1.6\pm 1.5$ &$-$  \\
 $a_{2}^{0}\pi^{0}$        &$13.5\pm 1.6\pm 3.2$ & $-$ &$-$  \\
 $a_{2}^{\pm}\pi^{\mp}$    &$-$ & {$18.4\pm 3.3\pm 5.5$} &$-$  \\
\hline\hline
\end{tabular}
\end{table}

The large branching fraction of $\chict\to K^{\ast}_{2}\overline{K}$ is a direct indication of the significant HSR violation effects. Note that the helicity amplitude ratios $|F_{2,0}|^2/|F_{1,0}|^2$, estimated with the fitted $g_{LS}$ (see Table~\ref{helratio}), suggest the dominance of $F_{1,0}$  in the transition amplitudes. The amplitude $F_{1,0}$ contributes to the leading HSR violation effects and scales as $(\Lambda_{\rm QCD})/ m_c)^6 $ due to its asymptotic behavior~\cite{Chernyak:1981zz,zhaoq}. In comparison with the HSR conserved channel $\chi_{c2}\to VV$, which scales as $(\Lambda_{\rm QCD}/ m_c)^4$, the ratio of $\chict\to K^{\ast}_{2}\overline{K}$ to $VV$ is expected to be suppressed by a factor of $(\Lambda_{\rm QCD}/ m_c)^2\sim 0.02$ with $\Lambda_{\rm QCD}\sim 0.2$~\gevcc and the charm quark mass $m_c\sim 1.5$~\gevcc. However,  the measured branching fraction of $\chict\to K^{\ast}_{2}\overline{K}$ appears to be the same order of magnitude as that for $\chi_{c2}\to VV$~\cite{pingrg}, which indicates a significant violation of HSR in $\chi_{c2}\to K^{\ast}_{2}\overline{K}$.

\begin{table}[hpbt]
\centering
\caption{The measured ratios of helicity amplitude squared $|F_{2,0}|^2/|F_{1,0}|^2$,
where the uncertainties are statistical only.}
\label{helratio}
\begin{tabular}{cccc}
 \hline\hline
 &~~~$\kk\pi^0$~~~ & \multicolumn{2}{c}{$\ks K^\pm \pi^\mp$}\\\cline{2-4}
 &~~ Charged $K^\ast$~~ &~Charged $K^\ast$~& ~Neutral $K^\ast$~ \\\hline
$K^{\ast}_2(1430)$ & $0.046\pm 0.001$ & $0.042\pm0.019$ & $0.031\pm0.018$   \\
  \hline\hline
\end{tabular}
\end{table}

In the analysis of the decay $\psip\to \gamma \pi^{+}\pi^{-}\pi^{0}$, the $\chi_{c2}$ signal is extracted by the requirement $|M_{\pi^{+}\pi^{-}\pi^{0}}-M_{\chi_{c2}}|\le 15$~\mevcc. The potential background from direct $e^{+}e^{-}$ annihilation is found to be  negligible by studying the continuum data taken at $\sqrt{s}=3.65$~GeV.
The backgrounds from $\psip$ decay are investigated with the $\psip$ inclusive MC sample; the only surviving $\chi_{c2}$ events are those that directly decay to $\pi^{+}\pi^{-}\pi^{0}$ without any intermediate state. There are also non-$\chi_{c2}$ backgrounds, which can be estimated by the events in the $\chi_{c2}$ sideband regions. Figure~\ref{pwa:gkskpi}(d) shows the invariant mass of $\pi^{\pm}\pi^{0}$ for the selected candidates, together with  the binned likelihood fit results. Here the fit components include the $\rho(770)^\pm$ signal, the direct decay $\chi_{c2}\to\pi^+\pi^-\pi^0$ and the non-$\chi_{c2}$ background.
The $\rho(770)^{\pm}$ signal and the direct $\chi_{c2}$ three-body decay are modeled with the MC simulated shapes convoluted with a Gaussian function with free parameters. The resonant parameters of the $\rho(770)^{\pm}$ are set to the values in the PDG~\cite{PDG}. The fitted signal yields are $14.7\pm 8.9$ and $63.6\pm 13.0$, and the corresponding resultant branching fractions are $(0.64\pm 0.39\pm 0.07)\times 10^{-5}$ and $(2.1\pm 0.4\pm 0.2)\times 10^{-5}$  for $\chi_{c2}\to\rho(770)^{\pm}\pi^{\mp}$ and the direct decay $\chi_{c2}\to\pi^{+}\pi^{-}\pi^{0}$, respectively, where the first uncertainties are statistical, and the second are systematic.
Since the statistical significance for the $\chi_{c2}\to\rho(770)^{\pm}\pi^{\mp}$ is only $2.8\sigma$, the upper limit at the 90\% C.L. for the branching fraction is set to $1.1\times10^{-5}$ by the method of Feldman-Cousins approach with the systematic uncertainties consideration~\cite{up}.

The uncertainties from the branching fractions of $\psip\to\gamma\chi_{c2}$, $\ks\to\pi^{+}\pi^{-}$, $\pi^{0}\to\gamma\gamma$, $a_{2}(1320)\to K\overline{K}$, and $K^{\ast}\to K\pi$ are quoted from the PDG~\cite{PDG}. The uncertainty on the number of $\psip$ events is about $0.8\%$~\cite{wang,wang1}. The uncertainties associated with the tracking and PID are 1\% for every charged track~\cite{xugf}. The uncertainty related with EMC shower reconstruction efficiency is 1\% per shower~\cite{xugf}. The uncertainties associated with the kinematic fit are estimated to be 0.5\% and 0.6\% for the 4C and 5C fit, respectively, by using a method to correct the charged-track helix parameters~\cite{guoyp}. The uncertainty associated with the $\ks$ reconstruction is estimated to be 2.5\%~\cite{guoyp}. The uncertainties related with the $\pi^{0}$ selection, the requirements on the $\pi^{0}$ recoil mass and the $\omega$ background veto (in $\chi_{c2}\to\rho(770)^{\pm}\pi^{\mp}$ mode only) are negligible.

In the decay $\chi_{c2}\to\rho(770)^{\pm}\pi^{\mp}$, the uncertainties due to the bin size and the fit range in the fit are estimated by repeating the fit with alternative bin sizes and fit ranges. The uncertainty due to the shape of $\chi_{c2}\to\pi^{+}\pi^{-}\pi^{0}$ is estimated by replacing the MC simulated line shape with a $3^{rd}$ polynomial function. The uncertainty due to the shape of the background is estimated by changing the $\chi_{c2}$ sideband regions.

In the decay $\chi_{c2}\to K\overline{K}\pi$, the uncertainties due to the contribution from $K^*(1410)$ and $K^*(1680)$ are estimated by including these states in the fit. The uncertainties associated with the backgrounds are determined by changing the $\chi_{c2}$ and $\pi^0$ sideband regions. The spin density matrix corresponding to the $E1$ transition~\cite{E1relation} is used in the nominal fit. To estimate the uncertainty, contributions from the quadrupole ($M2$) and other high order multipoles to the matrix~\cite{e1transition} are included in the fit, and the changes in the final results are treated as a systematic uncertainty. The uncertainties associated with the resonance parameters of intermediate states are estimated by varying their values by $1\sigma$ of their uncertainties quoted in the PDG~\cite{PDG}. The uncertainty due to the barrier radius \cite{zoubs:2003pj} when calculating $B_L(r)$ in Eq.~(\ref{pwa:f}) is estimated by alternative fits with $r=0.25$ or 0.75~fm, respectively, where $r=0.6$~fm is the nominal value.  The uncertainty associated with the direct three-body decay $\chi_{c2}\to K\overline{K}\pi$ is estimated by alternative fits with other spin-parity hypotheses, $e.g.$ a $0^{-}$ or $3^{-}$ non-resonant component in the $K\overline{K}$ or $K\pi$ systems. The largest changes in the signal yields are taken as systematic uncertainties. Assuming all the systematic errors are independent, the overall systematic obtained by taking the quadrature sum of the individual values.

In summary, the HSR suppressed processes of $\chi_{c2}\to K^{\ast}(892)\overline{K}$ and $\chi_{c2}\to \rho(770)^{\pm}\pi^{\mp}$ are studied with the $\psip$ data collected by BESIII for the first time.
The branching fractions of $\chi_{c2}\to K^{\ast}(892)^{\pm} K^{\mp}$ and $\chi_{c2}\to K^{\ast}(892)^{0}\overline{K}{}^{0}+\rm{c.c.}$ are measured to be $(1.5\pm 0.1\pm 0.2)\times 10^{-4}$ and $(1.3\pm 0.2\pm 0.2)\times 10^{-4}$, respectively, which are rather sizeable with respect to those of the HSR conserving decay $\chi_{c2}\to VV$~\cite{PDG,pingrg}. These branching fractions are at least one order of magnitude larger than the upper limit of the branching fraction of $\chi_{c2}\to \rho(770)^{\pm}\pi^{\mp}$ $(1.1\times 10^{-5})$. It is worth noting that this phenomenon is anticipated by the HSR violation mechanism proposed in Ref.~\cite{zhaoq}. Namely, the HSR violation in $\chi_{c2}\to K^{\ast}(892)\overline{K}$ occurs via the intermediate meson loops due to the large $U$-spin symmetry breaking, while that in $\chi_{c2}\to \rho(770)^{\pm}\pi^{\mp}$ is due to isospin symmetry breaking. Due to the large mass difference between $s$ and $u/d$ quarks, the $U$-spin symmetry is broken more severely in comparison with isospin symmetry. This results in the larger decay branching for $\chi_{c2}\to K^{\ast}(892)\overline{K}$ than that for $\chi_{c2}\to \rho(770)^{\pm}\pi^{\mp}$.  The results are crucial for further quantifying the HSR violation mechanisms~\cite{zhaoq} and also provide deeper insights into the underlying strong interaction dynamics in the charmonium energy region.

The BESIII collaboration thanks the staff of BEPCII and the IHEP computing center for their strong support. This work is supported in part by National Key Basic Research Program of China under Contract No. 2015CB856700; National Natural Science Foundation of China (NSFC) under Contracts Nos. 11175188, 11375205, 11425525, 11565006, 11521505, 11235011, 11322544, 11335008, 11425524, 11635010; the Chinese Academy of Sciences (CAS) Large-Scale Scientific Facility Program; the CAS Center for Excellence in Particle Physics (CCEPP); the Collaborative Innovation Center for Particles and Interactions (CICPI); Joint Large-Scale Scientific Facility Funds of the NSFC and CAS under Contracts Nos. U1232201, U1332201; CAS under Contracts Nos. KJCX2-YW-N29, KJCX2-YW-N45; 100 Talents Program of CAS; National 1000 Talents Program of China; INPAC and Shanghai Key Laboratory for Particle Physics and Cosmology; German Research Foundation DFG under Contracts Nos. Collaborative Research Center CRC 1044, FOR 2359; Istituto Nazionale di Fisica Nucleare, Italy; Joint Large-Scale Scientific Facility Funds of the NSFC and CAS	under Contract No. U1532257; Joint Large-Scale Scientific Facility Funds of the NSFC and CAS under Contract No. U1532258; Koninklijke Nederlandse Akademie van Wetenschappen (KNAW) under Contract No. 530-4CDP03; Ministry of Development of Turkey under Contract No. DPT2006K-120470; National Natural Science Foundation of China (NSFC) under Contract No. 11575133; NSFC under Contract No. 11275266; The Swedish Resarch Council; U. S. Department of Energy under Contracts Nos. DE-FG02-05ER41374, DE-SC-0010504, DE-SC0012069; U.S. National Science Foundation; University of Groningen (RuG) and the Helmholtzzentrum fuer Schwerionenforschung GmbH (GSI), Darmstadt; WCU Program of National Research Foundation of Korea under Contract No. R32-2008-000-10155-0.

\end{document}